\documentclass[twocolumn,showpacs,showkeys,superscriptaddress]{revtex4}
\usepackage{graphicx}
\usepackage{bm}
\usepackage{color}
\usepackage{amsmath}
\usepackage{amsfonts}
\usepackage{amssymb}
\usepackage{natbib}
\usepackage{latexsym}
\usepackage{mathrsfs}

\begin{document}

\title{Nondiffracting gravitational waves}

\author{Felipe A. Asenjo}
\email{felipe.asenjo@uai.cl}
\affiliation{Facultad de Ingenier\'ia y Ciencias,
Universidad Adolfo Ib\'a\~nez, Santiago 7491169, Chile.}
\author{Sergio A. Hojman}
\email{sergio.hojman@uai.cl}
\affiliation{Departamento de Ciencias, Facultad de Artes Liberales,
Universidad Adolfo Ib\'a\~nez, Santiago 7491169, Chile.}
\affiliation{Departamento de F\'{\i}sica, Facultad de Ciencias, Universidad de Chile,
Santiago 7800003, Chile.}
\affiliation{Centro de Recursos Educativos Avanzados,
CREA, Santiago 7500018, Chile.}

\begin{abstract}
It is proved that accelerating nondiffracting gravitational Airy wave--packets are solutions of linearized gravity. It is also showed that Airy functions are exact solutions to Einstein equations for non--accelerating nondiffracting gravitational wave--packets.
\end{abstract}


\maketitle
 
\section{Introduction}
 
Nondiffracting wave--packets are known solutions to Schr\"odinger equation \cite{berry,yuce} and Maxwell equations  \cite{rivka1,Hacyan}, and several experimental studies have been dedicated to them \cite{rivka1,Hacyan,ale,ionani,niko,niko2,rivka2,Bandres,Esat}. These solutions have been observed in experiments and some of their properties have been determined. Some of their features, such as acceleration, are due to the fact that they exhibit non--vanishing Bohm potential \cite{sahfaz201}. In fact, in Ref.~\cite{sahfaz201} it is established that  the usual dispersion relation for massless fields $k_\mu k^\mu=0$ is valid only when the Bohm potential associated to the solution vanishes, but otherwise the modified right-hand side is exactly equal to the Bohm potential, which may be either positive or negative. It is important to stress that similar results have been found by different authors in the last seventy years \cite{slep,skro,pleb,dewitt,vz1,mas1,har,mas2,mas3,sahthesis,hor,sah1,riv1,ah0,ah1,ah2,petrov,mugnai,gio,bouch,riv2,konda}, in addition to the ones already cited above.

Nevertheless, studying the implications of the possible existence of 
nondiffracting gravitational wave propagation seems 
to have been neglected so far. To understand its importance on gravitational wave behavior, research in this area is needed. 
This work intends to help filling that gap. 

This manuscript is devoted to investigate accelerating and non--accelerating nondiffracting gravitational waves. We show below how accelerating wave packets are solutions of linearized gravity. Additionally, we explicitly obtain exact non--accelerating,  nondiffracting gravitational wave solutions that, as far as we know, have not been considered before. This result consists on a new solution for the Ehlers-Kundt waves \cite{Ehlers,misner}.

\section{Accelerating linearized gravitational waves}

First, let us show that accelerating gravitational waves are solutions to linearized gravity.
As usual, let us consider a small perturbation $h_{\alpha\beta}$ of a background flat space--time metric $\eta_{\alpha\beta}=(-1,1,1,1)$, such that the total metric is $g_{\alpha\beta}=\eta_{\alpha\beta}+h_{\alpha\beta}$. It is straightforward to show that  the trace--reversed metric perturbation ${\overline{h}}_{\alpha\beta}=h_{\alpha\beta}-({1}/{2})\eta_{\alpha\beta}\, h_\mu^\mu$ \cite{misner},
satisfies the free wave equation
\begin{equation}
    \partial_\mu\partial^\mu   {\overline{h}}_{\alpha\beta}=0\, ,
    \label{eqevolution}
\end{equation}
under the Lorenz gauge
$\partial_\beta {\overline{h}}_{\alpha}^{\beta}=0$.

In order to exemplify the accelerating solutions for gravitational waves, let us focus in the evolution of the wave polarization ${\overline{h}}_{zz}={\overline{h}}_{zz}(t,x,y)$. Notice that $\partial_z {\overline{h}}_{zz}=0$,  thus satisfying the Lorenz gauge. Now we assume the following form  for such polarization 
\begin{equation}
    {\overline{h}}_{zz}(t,x,y)={\cal G}(\zeta,y)\exp\left(i k\, \eta\right)\, ,
    \label{fullsolution}
\end{equation}
where $k$ is an arbitrary constant,
$\zeta=x-t$ and $\eta=x+t$ \cite{bisieris}. Thus, as ${\overline{h}}_{zz}$ evolves according to Eq.~\eqref{eqevolution}, we are able to find  that ${\cal G}$ evolves as
\begin{equation}
    4ik\frac{\partial {\cal G}}{\partial\zeta}+\frac{\partial^2{\cal G}}{\partial y^2}=0\, .
\label{ecSchorGravwave}
\end{equation}
This equation is equivalent to a Schr\"odinger equation for a free particle. It is very well--known that this equation allows solution for accelerating wavepackets \cite{berry}. In our case, the accelerating gravitational wavepackets that solve Eq.~\eqref{ecSchorGravwave} are
\begin{equation}
    {\cal G}(\zeta,y)={\mbox{Ai}}\left(2 k\, y-k^2\zeta^2\right) \exp\left(2ik^2\zeta \, y-\frac{2i}{3}k^3\zeta^3 \right)\, ,
    \label{accGRAiry}
\end{equation}
where ${\mbox{Ai}}$
is an Airy function. Thus, the nondiffracting accelerating gravitational wavepacket is the real part of full solution \eqref{fullsolution}. This Airy wavepacket has an acceleration equal to $k/2$, deflecting  its trajectory in a parabolic path in the  $\zeta-y$ plane \cite{rivka1,Georgios}.  Airy waves have been extensively studied in the realm of light propagation in media \cite{bochen,cccchen,ChidaoChen,yupeng}. 
Beyond the 
theoretical importance of general accelerating wavepackets, which have encompassed several decades of research, the most remarkable fact is that they have been observed \cite{rivka2,Esat,sivi}.

Wave solution \eqref{accGRAiry}  has infinite energy, which is similar to what happens with  plane waves. However, structured nondiffracting  gravitational Airy wavepackets with finite energy can  be constructed
 using Gaussian beams
\cite{Georgios,Georgios2,Lekner}. The finite energy gravitational Airy wavepacket is given by
\begin{eqnarray}
&&    {\cal G}(\zeta,y)={\mbox{Ai}}\left(2 k\, y-k^2\zeta^2+ i a  k\zeta\right)\, \nonumber\\ && \quad\times \exp\left(a k y-a k^2 \zeta^2+\frac{i}{4} a^2 k \zeta+2ik^2\zeta \, y-\frac{2i}{3}k^3\zeta^3 \right)\, ,\nonumber\\
&&
    \label{accGRAiryFinite}
\end{eqnarray}
in terms of an arbitrary real parameter $a$ that makes possible the normalization of the solution. 
The expression given in \eqref{accGRAiryFinite} can be straightforwardly  proved to satisfy Eq.~\eqref{ecSchorGravwave}, and when $a=0$ we recover the wavepacket \eqref{accGRAiry}. 

We obtain that $\int dy\ |{\cal G}|^2=\int dy\ {\cal G}^*{\cal G}$, is independent of $\zeta$, by using Eq.~\eqref{ecSchorGravwave}. Then we get that this wavepacket is square integrable along $\zeta=0$ ($x=t$) as
\begin{equation}
\int_{-\infty}^\infty dy  |{\cal G}(\zeta,y)|^2=\int_{-\infty}^\infty dy |{\cal G}(0,y)|^2=\frac{e^{a^3/12}}{4k\sqrt{a \pi}}\, ,
\end{equation}
implying its finite energy.

\begin{figure}[h]
  \includegraphics[height=85mm]{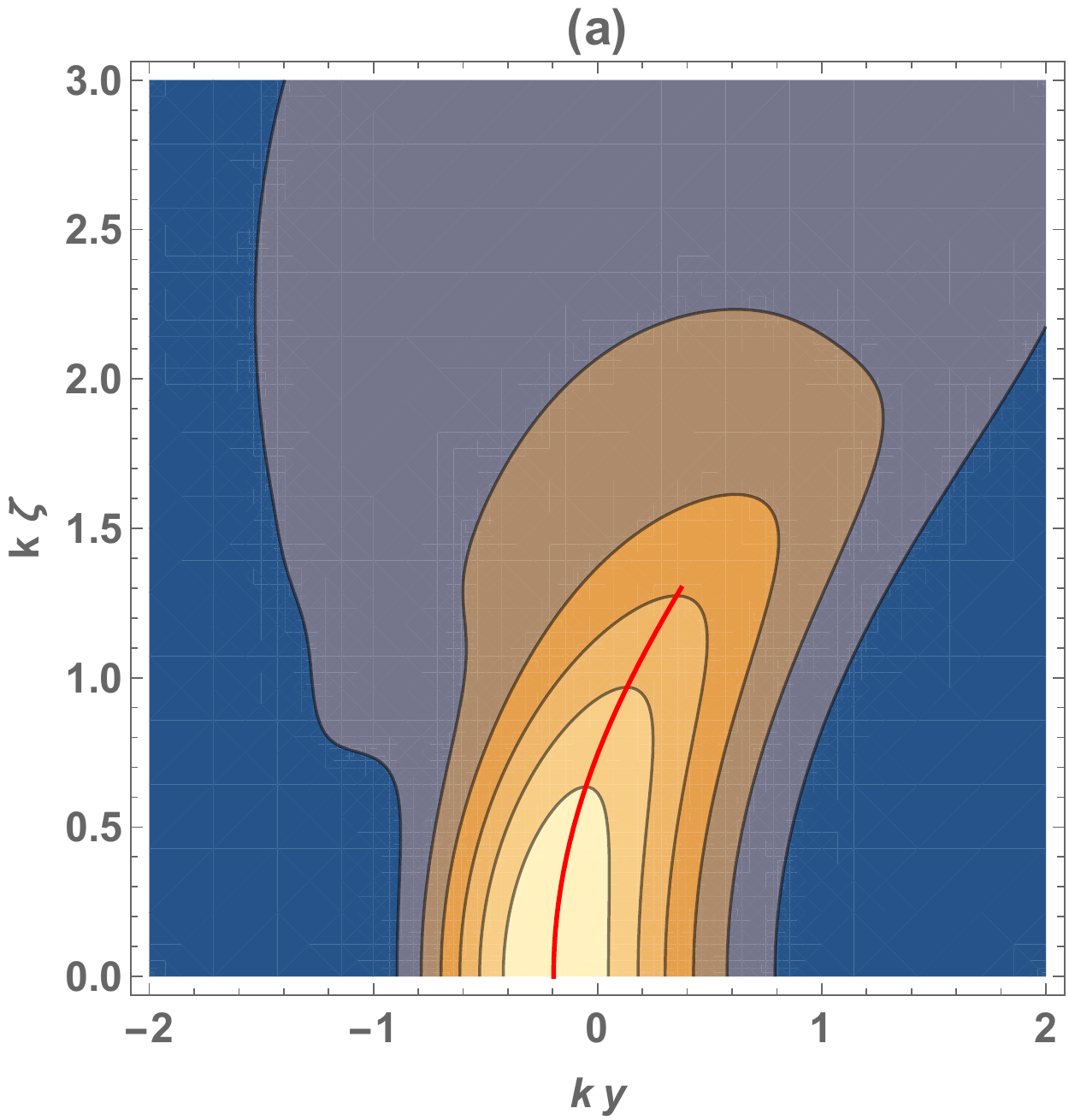}
  \\
      \includegraphics[height=85mm]{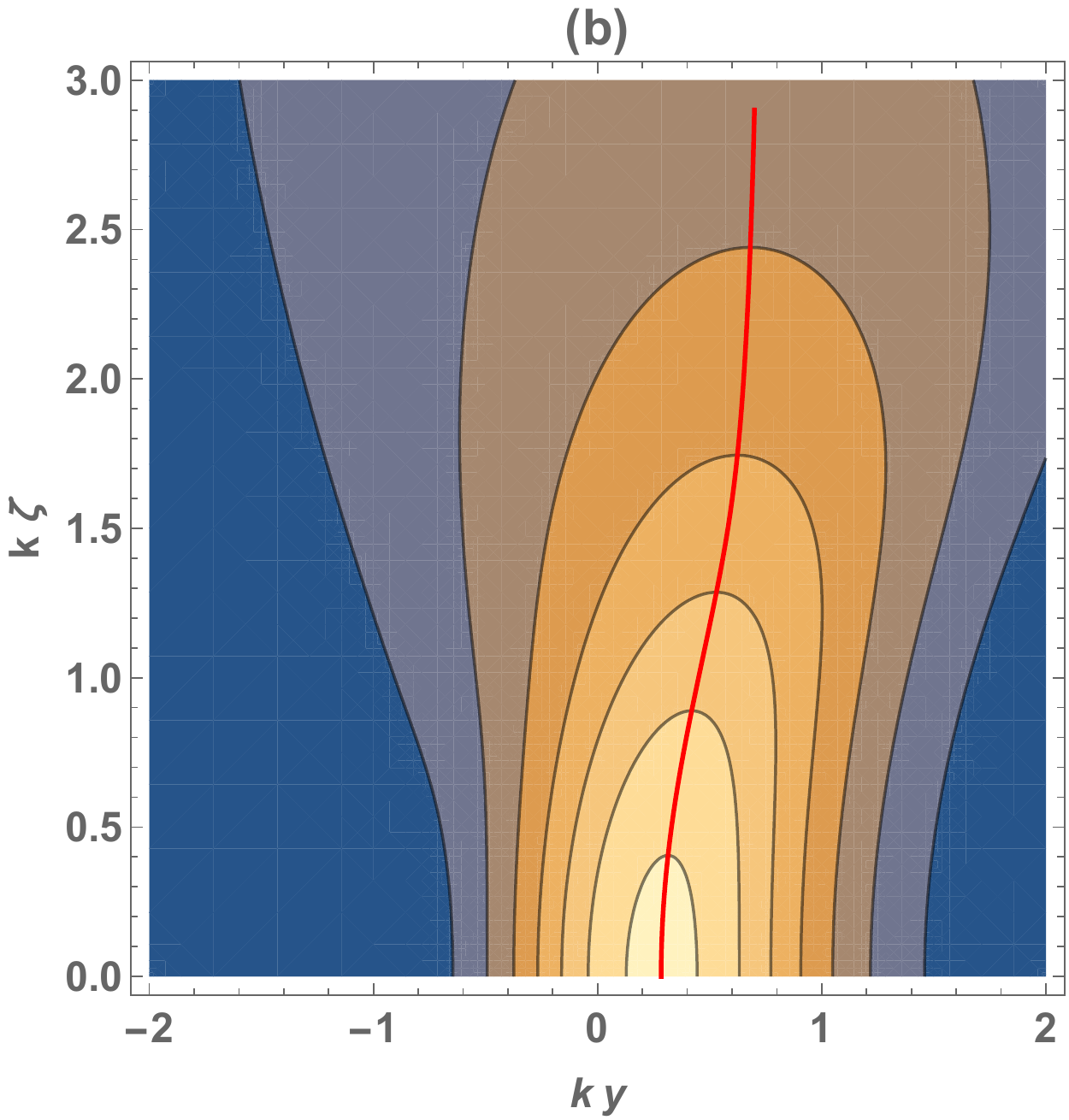} 
\caption{Contour plot for $|{\cal G}(\zeta,y)|$, with ${\cal G}$ given in Eq.~\eqref{accGRAiryFinite}, in terms of normalized $k\zeta$ and $k y$ variables, for different values of $a$.
(a) Plot for $a=1$. (b) Plot $a=2$. For both cases the red solid line shows the dynamical trajectory of the  position of local maxima
lobes of the wavepacket, which shows acceleration in the $\zeta-y$ plane. The colors get brighter as the value of $|{\cal G}(\zeta,y)|$ increases.}
\label{fig1final}. 
\end{figure}

As it is shown in Ref.~\cite{Lekner}, the gravitational 
Airy wavepackets \eqref{accGRAiryFinite}
do not accelerate. However, the main maxima lobe of the wave train does show
acceleration in the $\zeta-y$ plane. This is exemplified in Fig.~\ref{fig1final}, where a contour plot in  $\zeta-y$ plane of function
$|{\cal G}(\zeta,y)|=\sqrt{{\cal G}^*(\zeta,y){\cal G}(\zeta,y)}$
is shown for $a=1$ and $a=2$. In red solid lines we display the evolution of the position of the maxima lobe, that moves in a curved trajectory, thus displaying its acceleration in such plane.

\section{Exact nondiffracting Gravitational waves}

Structured  wave--packets are not only solutions to the linearized Einstein equations, but they are also exact solutions to the full Einstein equations. Exact nondiffracting
 gravitational waves can be obtained from a modification of Ehlers--Kundt solution \cite{Ehlers,misner}. Consider the exact metric in cartesian coordinates given by the interval
\begin{eqnarray}
ds^2=-dt^2+ L(u)^2dx^2+W(u)^2dy^2+dz^2\, ,
\label{metricexactpp}
\end{eqnarray}  
where  $L$ and $W$ are arbitrary functions of $u=z-t$. 
Einstein equations are satisfied when
\begin{eqnarray}
\frac{1}{L}\frac{d^2 L}{du^2}+\frac{1}{W}\frac{d^2 W}{du^2}=0\, .
\label{eqExacrGW}
\end{eqnarray}
These waves, described by metric \eqref{metricexactpp}, are {\it pp}-waves \cite{misner,griffiths}. This can be seen by applying the coordinate transformation $x=X/L$, $y=Y/W$, $v=V-X^2 d_u L/L-Y^2 d_uW/W$, together with Eq.~\eqref{eqExacrGW}, to put the metric \eqref{metricexactpp} in the Kerr--Schild form
$ds^2=dX^2+dY^2-du\, dV+du^2(X^2-Y^2)d_u^2 L/L$.  For these classes of metrics
linear and non--linear theories coincide under certain
conditions \cite{Kilicarslan}.

Here we show that there are certain solutions of  Eq.~\eqref{eqExacrGW} that
describe exact nondiffracting and non--accelerating  gravitational waves. 
The simplest  non--trivial solution for a structured gravitational wave can be obtained in terms of Airy functions
\begin{eqnarray}
L(u)&=&L_0\, {\mbox{Ai}}(u)\, ,\nonumber\\
W(u)&=&W_0\, {\mbox{Ai}}(-u)\, ,
\label{eqcaxsGrAiy}
\end{eqnarray}
with arbitrary constants $L_0$ and $W_0$.
These nondiffracting wave--packets have the property that the intensity of the gravitational wave is concentrated in the caustic. 
However, these exact Airy gravitational waves have infinite energy.
This is a feature that they share with other gravitational waves, such plane waves \cite{misner}. Nevertheless, what is relevant in this exact solution is its structured configuration.

\section{Discussion}

Nondiffracting  wave--packets have been a fruitful field of study in quantum mechanics \cite{berry,yuce,dirac,nutevolo}, electromagnetism and optics \cite{rivka1,Hacyan,ale, ionani,niko,niko2,rivka2,Bandres,Esat}. 
The solutions \eqref{accGRAiry} and \eqref{eqcaxsGrAiy} presented along this work are in the same spirit, and complement others solutions for structured gravitational beams, as Bessel or Laguerre-Gauss, found in linearized general relativity \cite{iwobyu,patak}.

Although simple in their mathematical form, the new linearized accelerating and exact non--accelerating solutions can bring new insights in the form that different gravitational waves propagate in a flat space--time background. 

How the nondiffracting features are manifested for gravitational
waves in curved space--time is an open question currently under investigation.


\end{document}